\documentclass[12pt]{article}

\usepackage[dvips]{epsfig}
\usepackage{graphicx}
\usepackage{latexsym}
\usepackage{verbatim}
\usepackage{amsmath}
\usepackage{amsthm}
\usepackage{amssymb}
\usepackage{a4wide}
\usepackage{subfigure}

\theoremstyle{plain}

\theoremstyle{remark}

\numberwithin{equation}{section}

\newcommand{\R}{\mathbb{R}}

\newcommand{\eps}{\varepsilon}

\newcommand{\order}{{\cal O}}

\newcommand{\del}{\partial}
\newcommand{\X}{Z}
\newcommand{\V}{{S^2}}

\title{Adaptive Finite Element Simulation of the Time-dependent Simplified $P_N$ Equations}

\author{
Martin Frank\thanks{TU Kaiserslautern, Fachbereich Mathematik, Erwin-Schr\"odinger-Str., 67663 Kaiserslautern, Germany, {\tt frank@mathematik.uni-kl.de}}
\and
Jens Lang\thanks{TU Darmstadt, Fachbereich Mathematik, Schlossgartenstrasse 7, 64289 Darmstadt, Germany, {\tt lang@mathematik.tu-darmstadt.de}}
\and 
Matthias Sch\"afer\thanks{Fraunhofer ITWM, Fraunhoferplatz 1, 67663 Kaiserslautern, Germany, {\tt matthias.schaefer@itwm.fhg.de}} 
}

\date{January 20, 2009}

\begin{document}

\maketitle

\begin{abstract}
The steady-state simplified $P_N$ approximation to the radiative transport equation has been successfully applied to many problems involving radiation. Recently, time-dependent simplified $P_N$ equations have been derived by an asymptotic analysis similar to the asymptotic derivation of the steady-state $SP_N$ equations \cite{FraKlaLarYas07}. In this paper, we present computational results for the time-dependent $SP_N$ equations in two dimensions, obtained by using an adaptive finite element approach. Several numerical comparisons with other existing models are shown.
\end{abstract}

\noindent

\section{Introduction}
%
Time-dependent radiative transfer, described by the radiative transfer equation, is hard to compute. This is due to the six-dimensional phase space (1$\times$ time, 2$\times$ angle, 3$\times$ space). There is an interest in time-dependent radiative transfer solutions, e.g.\ in astrophysics (supernova explosions),
the interaction of short-pulsed lasers with plasmas, and light detection and ranging (LIDAR). It is our purpose in this paper to numerically investigate the time-dependent simplified $P_N$ equations in two dimensions. These models have been very successful in the steady case. Here we investigate an extension to the time-dependent case.

The simplified $P_N$ ($SP_N$) equations were originally developed for steady-state problems in nuclear engineering \cite{Gel60,Gel61,Gel62} and have subsequently been generalized and successfully applied in several other fields, including radiative transfer \cite{KlarLangSeaid2005,LarThoKla02,LarThoKlaSeaGot02}. The first formal derivation by Gelbard \cite{Gel60,Gel61,Gel62} started with the one-dimensional $P_N$ equations, which contain only first-order space derivatives, and used substitutions to obtain a system of elliptic partial differential equations. To obtain equations in three space dimensions, even-order moments are interpreted as scalars, odd-order moments are interpreted as vectors, and one-dimensional derivatives
$\partial_x$ are replaced by divergence operators and gradients respectively. In three space dimensions, compared to the $(N+1)^2$ independent unknowns in the spherical harmonics $P_N$ equations, the number of unknowns in the $SP_N$ equations increases only linearly with $N$. Because of the derivation via the one-dimensional $P_N$ equations, the $SP_N$ method was at first not widely accepted. But alternative derivations via asymptotic expansion \cite{LarMorMcG96} and via a variational approach \cite{BraLar00,TomLar96} have substantiated the validity of the $SP_N$ hierarchy.

The $SP_N$ equations are accurate if the medium is optically thick, the scattering rate is comparable to the collision rate, and scattering is not highly forward-peaked \cite{LarMorMcG96}. In addition, numerical experiments (cf.\ \cite{LarThoKlaSeaGot02} and references therein) have shown that the $SP_N$ equations give good results even when the regime is not so diffusive, and even in the presence of a discontinuity in the opacities. This means that in the diffusive regime a higher accuracy is obtained and at the same time the range of applicability is increased.

Until recently, the $SP_N$ method was almost exclusively applied to steady-state transport equations, i.e.\ no time dependence was assumed. Only then can the $P_N$ equations be substituted into each other to give a second-order system. To our knowledge, there has been only one attempt in the literature \cite{MorMcGLar96} to apply the $SP_N$ method to a time-dependent problem. Here, the authors use a semi-discretization in time (i.e.\ the time variable is discretized whereas the other variables are treated as being continuous) and apply the $SP_N$ approximation to the then steady system. This paper, on the other hand, investigates time-dependent $SP_N$ equations which were systematically derived from the Boltzmann equation using an asymptotic analysis.

The numerical solution of the $SP_N$ equations are often still quite expensive 
due to their inherent multi-scale structure in both time and space. A remedy is to
use fully adaptive algorithms where the local accuracy of the
numerical solution is controlled by means of a posteriori error estimates
in space and time. Such estimators are well established to control the adaptive multilevel 
process producing highly refined space-time grids to capture local effects efficiently and 
therefore drastically reducing the size of the arising linear algebraic systems with respect to a
prescribed tolerance. We apply the adaptive Rothe method based
on the discretization sequence first in time then in space, in contrast to the
usual Method of Lines approach (see e.g. \cite{Lang2000} and references therein). The spatial
discretization is considered as a perturbation of the time integration process.
Implementations have been done in the {\sf KARDOS} library \cite{ErdmannLangRoitzsch2002}, which
provides a suitable programming environment for adaptive algorithms to solve nonlinear time-dependent PDEs.

This paper is organized as follows: A brief summary of the derivation of the time-dependent $SP_N$ equations using asymptotic analysis is given in Section \ref{sec:TDSPN}. Suitable initial and boundary conditions are stated in Section \ref{sec:BC}. The imployed numerical method is described in Section \ref{sec:NUM}. In Section \ref{sec:SIM}, these techniques are applied to two test cases from the recent radiative transfer literature.

\section{Time-Dependent $SP_N$ Equations}
\label{sec:TDSPN}
%
We consider a convex, open, bounded domain $Z$ in $\R^3$, and we assume that $Z$ has a smooth boundary with outward normal vector $n$. The direction of particle motion is given by $\Omega\in S^2$, where $S^2$ is the unit sphere in three dimensions. Moreover, we let
$$
\Gamma = \partial \X\times \V \quad\text{and}\quad \Gamma^- = \{ (x,\Omega)\in \Gamma: n(x) \cdot \Omega < 0\}.
$$

The transport of mono-energetic particles that undergo isotropic scattering in a medium is modeled by the linear Boltzmann equation
\begin{equation}\label{pde}
 	 \frac{1}{v}\partial_t \psi(t,x,\Omega) + \Omega \cdot \nabla_x \psi(t,x,\Omega) + \sigma_t(x)\psi(t,x,\Omega)
 	=  \frac{\sigma_s(x)}{4\pi}\int_\V \psi(t,x,\Omega') d\Omega' + \frac{q(t,x)}{4\pi},
\end{equation}
where $q$ is an isotropic source term.
At the boundary, we prescribe the ingoing radiation
\begin{equation}\label{eq:BC}
\psi(t,x,\Omega) = \psi_b(t,x,\Omega) \quad\text{on}\quad \Gamma^-,
\end{equation}
and as the initial condition, we prescribe
\begin{equation}\label{eq:IC}
\psi(0,x,\Omega) = \psi_0(x,\Omega).
\end{equation}
Here, $\psi(t,x,\Omega)\cos\theta dA dt d\Omega$ is the number of particles at point $x$ and time $t$ that move with velocity $v$ during $dt$ through an area $dA$ into a solid angle $d\Omega$ around $\Omega$, and $\theta$ is the angle between $\Omega$ and $dA$. The total cross section $\sigma_t(x)$ is the sum of the absorption cross section $\sigma_a(x)$ and the total scattering cross section $\sigma_s(x)$.

The time-dependent $SP_N$ equations have been derived in \cite{FraKlaLarYas07}. For the convenience of the reader, we here present an abbreviatied version which contains the major ideas.
The steady-state diffusion equation is an elliptic PDE. Time-dependent diffusion theory is governed by a parabolic PDE. To obtain higher-order corrections to diffusion theory, we write the transport equation in a parabolic scaling. Space-derivatives are scaled by a small parameter $\eps$ and the additional time-derivative is scaled by $\eps^2$. This is called a parabolic scaling, since a differential operator that is first-order in time and second-order in space is invariant under this scaling. The transport equation is therefore written as:
\begin{equation}\label{e1}
	\eps^2\frac{1}{v}\del_t \psi + \eps \Omega\cdot\nabla_x \psi + \sigma_t \psi = \left( \sigma_t -\eps^2 \sigma_a \right) \frac{1}{4\pi}\phi+ \eps^2 \frac{q}{4\pi},
\end{equation}
where $\psi=\psi(t,x,\Omega)$, $\phi(t,x) = \int_{S^2}\psi(t,x,\Omega)d\Omega$, and $q=q(t,x)$.

Integrating (\ref{e1}) over $\Omega$ and dividing by $\eps^2$, we obtain the ``balance" equation
\begin{equation}\label{eq:bal}
	\frac{1}{v}\del_t \phi + \frac{1}{\eps} \nabla_x\cdot \int_{S^2}\Omega \psi d\Omega + \sigma_a \phi = q,
\end{equation}
which states a basic physical principle:\ changes in the scalar flux $\phi$ are either due to leakage (the spatial derivative term), absorption, or sources.
We require that this ``balance'' equation be contained in the final choice of $SP_N$ equations.

We write (\ref{e1}) as
\begin{equation}\label{e2}
(1+\eps \Omega\cdot X+\eps^2 T) \psi = S,
\end{equation}
where
\begin{equation}\label{e3}
	X=\frac{1}{\sigma_t}\nabla_x,\quad T = \frac{1}{v\sigma_t}\del_t,\quad\text{and}\quad S=\left( 1-\eps^2\frac{\sigma_a}{\sigma_t}\right)\frac{\phi}{4\pi}+\eps^2\frac{q}{4\pi\sigma_t}.
\end{equation}

We start by expanding the inverse of the operator in (\ref{e2}) in powers of $\eps$
\begin{align}\label{e4}
\psi&=(1+\eps\Omega\cdot X +\eps^2 T)^{-1} S
\nonumber\\
&= \left\{
1-(\Omega\cdot X)\eps
+\left[-T+(\Omega\cdot X)^2\right]\eps^2
+\left[(\Omega\cdot X)T+(T-(\Omega\cdot X)^2)(\Omega\cdot X)\right]\eps^3
\right.
\nonumber\\
&\left.\quad
+\left[(T-(\Omega\cdot X)^2)T+(-2(\Omega\cdot X)T+(\Omega\cdot X)^3)(\Omega\cdot X)\right]\eps^4
\cdots
\right\}S + \order(\eps^5).
\end{align}
In the following we assume that the system is homogeneous, i.e.\ $\sigma_a$ and $\sigma_t$ are constant. This assumption is crucial for the validity of the following analysis. For a discussion of the non-homogeneous case, we refer the reader to the end of this section. Integrating (\ref{e4}) with respect to $\Omega$ and using
\begin{equation}\label{e5}
\int_{S^2} (\Omega\cdot X)^n d\,\Omega
=[1+(-1)^n]\frac{2\pi}{n+1} X^n = [1+(-1)^n]\frac{2\pi}{n+1} (X\cdot X)^\frac{n}{2},
\end{equation}
we obtain
\begin{align}\label{e6}
\phi=&\int_{S_2} \psi d\,\Omega
\nonumber\\
=&4\pi\left\{
1
+\left(\frac{1}{3}X^2-T\right)\eps^2
+\left(T^2+\frac{1}{5}X^4-TX^2\right)\eps^4
\right.
\nonumber\\
&\left.
+\left(\frac{1}{7}X^6+2T^2X^2-T^3-TX^4\right)\eps^6
\right\}S+\order (\eps^8).
\end{align}
Hence,
\begin{align}\label{e6a}
4\pi S=
&\left\{
1
+\left(\frac{1}{3}X^2-T\right)\eps^2
+\left(T^2+\frac{1}{5}X^4-TX^2\right)\eps^4
\right.
\nonumber\\
&\left.
+\left(\frac{1}{7}X^6+2T^2X^2-T^3-TX^4\right)\eps^6
\right\}^{-1}\phi+\order (\eps^8)
\nonumber\\
=&
\left\{
1
+\left(-\frac{1}{3}X^2+T\right)\eps^2
+\left(-\frac{4}{45}X^4+\frac{1}{3}TX^2\right)\eps^4
\right.
\nonumber\\
&\left.
+\left(-\frac{44}{945}X^6-\frac{1}{3}T^2X^2+\frac{4}{15}TX^4\right)\eps^6
\right\}\phi+\order (\eps^8).
\end{align}
Inserting the definition of the source term $S$ from (\ref{e3}), we get
\begin{equation}
\begin{split}
\left( 1-\eps^2\frac{\sigma_a}{\sigma_t}\right)\phi + \eps^2\frac{q}{\sigma_t}
=&
\left\{
1
+\left(-\frac{1}{3}X^2+T\right)\eps^2
+\left(-\frac{4}{45}X^4+\frac{1}{3}TX^2\right)\eps^4
\right.\\
&\left.
+\left(-\frac{44}{945}X^6-\frac{1}{3}T^2X^2+\frac{4}{15}TX^4\right)\eps^6
\right\}\phi+\order (\eps^8).
\end{split}
\end{equation}
Deleting $\phi$ on both sides and multiplying by $\sigma_t/\eps^2$, we obtain
\begin{equation}\label{e7}
\begin{split}
-\sigma_a\phi + q
= \sigma_t T \phi - \frac{\sigma_t}{3} X^2 \biggl[ &\phi - \varepsilon^2 T \phi
        + \frac{4}{15} \varepsilon^2 X^2 \phi  \\
     & + \frac{44}{315} \varepsilon^4 X^4 \phi + \varepsilon^4 T^2 \phi
        - \frac{4}{5} \varepsilon^4 T X^2 \phi \biggr] + \order(\varepsilon^6).
\end{split}
\end{equation}
We note that this equation has the form of the balance equation (\ref{eq:bal}). Since we want to keep this form, in the subsequent approximations we only manipulate the terms within the brackets.

\subsection{$SP_1$ Approximation}
%
For the lowest-order approximation, we neglect terms of order $\order(\eps^2)$. Then (\ref{e7}) becomes the classical diffusion ($SP_1$) equation
\begin{equation}
	\frac{1}{v}\partial_t \phi = \frac{1}{3\sigma_t}\nabla_x^2\phi-\sigma_a\phi+q.
\end{equation}


\subsection{$SP_3$ Approximation}
%
As in the steady case, the $SP_2$ equations, which are of order $\order(\eps^4)$, have proven to be inadequate in practice. This is due to their origin from the $P_2$ equations \cite{FraKlaLarYas07}. Therefore we omit them and proceed with the $SP_3$ equations.

Noting that Eq.\ (\ref{e7}) has the form of the balance equation (\ref{eq:bal}), we write (\ref{e7}) as
\begin{equation}
\begin{split}
q - \sigma_a\phi = \sigma_t T \phi -\frac{\sigma_t}{3} X^2\Big\{ &\phi + \Big[1+\frac{11}{21}\eps^2X^2-3\alpha \eps^2 T\Big]\frac{4}{15}\eps^2X^2\phi \\
&- \Big[1-\eps^2T+\frac{4}{5}(1-\alpha)\eps^2X^2\Big]\eps^2T\phi \Big\} + \order(\eps^6).
\end{split}\label{eq7}
\end{equation}
As before, we have isolated terms that contain time-dependent diffusion operators (first-order time and second-order space derivative). The transformation of the asymptotic expansion into the $SP_2$ system, i.e.\ the definition of $\xi$, is unique up to a multiplicative factor. However, for the expansion up to terms of order $\order(\eps^6)$, it is not clear how the substitutions have to be performed. Thus we have introduced a parameter $\alpha\in[0,1]$ to split the mixed term $TX^2$ into two parts. We chose the parameter between zero and one in order to get diffusion equations with the correct signs.

Using Neumann's series, we write (\ref{eq7}) as:
\begin{equation}\label{eq8}
\begin{split}
q - \sigma_a\phi = \sigma_t T \phi - \frac{\sigma_t}{3} X^2\Big\{ &\phi + \Big[1-\frac{11}{21}\eps^2X^2+3\alpha \eps^2 T\Big]^{-1}\frac{4}{15}\eps^2X^2\phi \\
&- \Big[1+\eps^2T-\frac{4}{5}(1-\alpha)\eps^2X^2\Big]^{-1}\eps^2T\phi \Big\} + \order(\eps^6).
\end{split}
\end{equation}
Now we define
\begin{subequations}\label{e9}
\begin{equation}
\phi_2=\frac{1}{2}\left[1-\frac{11}{21}\eps^2X^2+3\alpha\eps^2T\right]^{-1}\left(\frac{4}{15}\eps^2X^2\phi\right),
\end{equation}
\begin{equation}\label{e11}
\zeta=\left[1+\eps^2 T-\frac{4}{5}(1-\alpha)\eps^2X^2\right]^{-1}
\left(\eps^2T\phi\right),
\end{equation}
\end{subequations}
to obtain the system
\begin{subequations}\label{e13}
\begin{align}
\frac{1}{v}\partial_t\phi&=
\frac{1}{3\sigma_t}\nabla_x^2\left[ \phi+2\phi_2-\zeta \right]-\sigma_a\phi+q, \label{e12a}\\
\frac{1}{v}\partial_t\phi_2 & =\frac{1}{3\sigma_t}\nabla_x^2\left[\frac{2}{15\alpha}\phi+\frac{11}{21\alpha}\phi_2\right]-\frac{1}{3\alpha}\frac{\sigma_t}{\eps^2}\phi_2,\label{e12b}\\
\frac{1}{v}\partial_t\zeta&=\frac{1}{3\sigma_t}\nabla_x^2\left[\phi+2\phi_2+\left(\frac{12}{5}(1-\alpha)-1\right)\zeta\right]-\sigma_a\phi+q-\frac{\sigma_t}{\eps^2}\zeta.
\end{align}
\end{subequations}
Without time-dependence, the variable $\zeta$ is zero. Moreover, for $\alpha=\frac{2}{3}$ the above equations reduce to the steady-state $SP_3$ approximation. To obtain a system that is not ill-posed, we must take $0<\alpha<0.9$ \cite{FraKlaLarYas07}.

\subsection{Simplification of the $SP_3$ System}
\label{sec:STAB}
%
In \cite{FraKlaLarYas07}, the $SP_3$ equations with $\alpha=2/3$ were derived from the $P_3$ moment equations. The variable $\phi_2$ can be identified with the second-order Legendre moment of the radiative intensity. The variable $\zeta$, on the other hand, is an auxiliary variable without a straight-forward physical interpretation. Furthermore, $\zeta=0$ in steady-state.
To simplify the $SP_3$ equations, we therefore make a quasi-steady approximation and neglect $\zeta$. We obtain
\begin{subequations}\label{e18}
\begin{align}
\frac{1}{v}\partial_t\phi&=
\frac{1}{3\sigma_t}\nabla_x^2[\phi+2\phi_2] -\sigma_a \phi + q,
\label{e18a}\\
\frac{1}{v}\partial_t\phi_2&=
\frac{1}{3\sigma_t}\nabla_x^2\left[\frac{2}{15\alpha}\phi+\frac{11}{21\alpha}\phi_2\right]
-\frac{1}{3\alpha}\frac{\sigma_t}{\eps^2}\phi_2.
\label{e18b}
\end{align}
\end{subequations}
We call these the $SSP_3$ (simplified-simplified $P_3$) equations.

We expect that the time-dependent $SP_N$ equations can be generalized to anisotropic scattering in a similar manner as in the steady-state case \cite{LarMorMcG96}. In the derivation of the equations, we assumed a homogeneous medium. In steady-state, a variational analysis yielded the $SP_N$ equations for non-homogeneous media as well as interface and boundary conditions \cite{BraLar00,TomLar96}. The only difference for space-dependent coefficients is that the spatial derivatives have to be modified like
$$
\frac{1}{\sigma_t}\nabla_x^2 \quad\rightarrow\quad \nabla_x\frac{1}{\sigma_t(x)}\nabla_x.
$$
For steady-state problems, this modification of the spatial derivatives is asymptotically
correct in planar geometry and we expect that it is asymptotically correct for time-dependent planar geometry problems.

\section{Boundary Conditions and Initial Values}
\label{sec:BC}
%
In this section, we state boundary conditions for the $SP_N$ equations which have been derived using Marshak's method \cite{Mar47}. Let
\begin{equation}
l_1=-4\int_{n\cdot\Omega <0} (n\cdot\Omega)\psi_b\,d\Omega,\quad
l_2=16\int_{n\cdot\Omega <0} P_3(n\cdot\Omega)\psi_b\,d\Omega.
\nonumber
\end{equation}
For the $SP_1$ equations, we have
\begin{equation}\label{sp1_2}
n\cdot \nabla_x\phi = \frac{\sigma_t}{\eps} \left(\frac{3}{2}l_1 - \frac{3}{2}\phi\right).
\end{equation}
For $SP_3$ and $SSP_3$, we obtain the boundary conditions:
\begin{subequations}
\begin{align}
n\cdot\nabla_x \phi &=  \frac{\sigma_t}{\eps} \left(-\frac{25}{12}\phi+\frac{25}{24}\phi_2+\frac{3}{2}l_1+\frac{7}{12}l_2\right) \\
n\cdot\nabla_x \phi_2 &=  \frac{\sigma_t}{\eps} \left(\frac{7}{24}\phi-\frac{35}{24}\phi_2-\frac{7}{24}l_2\right) \\
\zeta &= 0.
\end{align}
\end{subequations}
\subsection{Initial Values}
Given an initial particle distribution, it is straight-forward to calculate an initial value for $\phi$. From the asymptotic analysis, the physical meaning of the auxiliary variables ($\xi$, $\phi_2$, $\zeta$) is not obvious. Therefore it is not clear what the appropriate initial conditions for these variables are. In many cases, the initial setting is a steady state. In addition, the time-dependent $SP_N$ equations reduce to the steady-state $SP_N$ equations.
For the $SP_3$ equations, we would have to solve (\ref{e9}) for $\phi_2$ and $\zeta$. Of course, this gives $\zeta=0$. Alternatively, $\phi_2$ could be identified as the second-order Legendre moment and thus be computed from the initial value for $\psi$.

\section{Numerical Method}
\label{sec:NUM}
We have derived the time-dependent $SP_N$ equations in three spatial dimensions. In the following chapter, we will present numerical results in two spatial dimensions. Therefore, here we also describe the numerical method for two dimensions. It can be generalized, however, in a straightforward way.

Mathematically speaking, the above $SP_N$ models are nonlinear
parabolic PDEs. This means that after spatial discretization we are faced with a
large scale stiff system. From an ODE point of view, an optimal treatment of this
stiffness structure is to apply some $L$-stable implicit
time discretization \cite{DeuflhardBornemann2002,HairerWanner1996}. From the PDE point of view,
the avoidance of order reduction (which may occur above order $2$) is equally important
for the overall efficiency of the time integrator. Both properties are
satisfied by the linearly implicit time discretization of Rosenbrock type
behind the code {\sf ROS3PL} \cite{LangTeleaga2008}.
Note that the popular Crank--Nicolson scheme is not $L$-stable and even not strongly
$A$-stable, which results in an insufficient filtering
of spurious modes. Fully implicit schemes require the iterative
solution of finite dimensional nonlinear systems of algebraic equations by some
Newton-like method. In contrast, linearly implicit methods realize a simplified Newton
method in function space and require only the solution of a fixed number of linear systems
per time step. According to their one-step nature, they allow
for a rapid change of step sizes and an efficient adaptation of
the spatial discretization in each time step. Moreover, a simple
embedding technique can be used to estimate the error part arising
from time discretization.

Linearly implicit time integrators of Rosenbrock type are implemented in
the code family {\sf KARDOS} \cite{ErdmannLangRoitzsch2002}, which is used
to solve our $SP_N$ models.
{\sf KARDOS} is characterized by a combination of Rosenbrock solvers in time with multilevel
finite elements in space in the setting of an adaptive Rothe approach, i.e., first
time discretization and then spatial discretization. In this setting, both
time-step control and dynamic mesh refinement on the basis of a posteriori error estimation can be
simultaneously realized \cite{Bornemann1992,LangWalter1992}. A rigorous analysis for nonlinear parabolic
systems has been given in \cite{Lang2000}, where challenging examples from other fields
of science and technology are also included.

Next we want to describe the main ingredients of the adaptive Rothe
method as needed for the efficient solution of the above described $SP_N$ models. These models
can be written as abstract Cauchy problems of the form
\begin{equation}
\label{lang_lirk_cauchy}
H\partial_tU = F(U),\quad U(t_0)=U_0,\quad t>0,
\end{equation}
where $H$ is a constant regular matrix and the diffusion operators and the boundary conditions are incorporated into the nonlinear function $F(U)$. For example, we have $U=(\phi,\xi,\zeta)^T$ for the $SP_3$ model. To approximate the vector $U(x,t)$ by values $U_n\approx U(\cdot,t_n)$ at a certain time grid
\begin{equation} \label{lang_timegrid} %
0 = t_0 < t_1 < \cdots < t_n < \cdots < t_{M-1} < t_M = T\,, %
\end{equation} %
we apply the $4$-stage third-order Rosenbrock method {\sf ROS3PL}, which has the recursive form
\begin{eqnarray}
\label{lang_lirk_rosen1}
\displaystyle \Bigg(\frac{H}{\tau_n\gamma}-\,J_n\Bigg)\,U_{ni} \!\!&\;=\;& \!\! \displaystyle
F\Bigg(U_n\!+\!\sum\limits_{j=1}^{i-1}
a_{ij}\,U_{nj}\Bigg) - H\sum\limits_{j=1}^{i-1}
\frac{c_{ij}}{\tau_n}\,U_{nj},
\quad i=1,\ldots,4,\\[2mm]
\label{lang_lirk_rosen2}
U_{n+1} \!\! &\;=\;&\!\! U_n + \sum\limits_{i=1}^{4}m_iU_{ni},
\end{eqnarray}
where $\tau_n=t_{n+1}-t_n$ and $J_n=F'(U_n)$.
The defining formula coefficients $m_i$, $a_{ij}$, $c_{ij}$, and $\gamma$ are
given in \cite{LangTeleaga2008}. The method is L-stable and avoids order reduction.

{\sf ROS3PL} offers a simple way to estimate the local error. An embedded solution
$\hat{U}_{n+1}$ of second order can be computed by replacing
the original weights $m_i$ by $\hat{m}_i$ in (\ref{lang_lirk_rosen2}). In order to take
into account the scale of the problem, the local error estimator
is defined by the weighted root mean square norm
\begin{equation}\label{lang_time_estim} r_{n+1}=
\left( \frac{\|U_{n+1}-\hat{U}_{n+1}\|^2_{L^2(Z)}}
{ATOL+RTOL\,\|U_{n+1}\|^2_{L^2(Z)}} \right)
^{1/2}.
\end{equation}
The tolerances $ATOL$ and $RTOL$ have to be selected carefully to furnish
meaningful input for the error control. The estimator
can be used to propose a new time step by
\begin{equation}\label{lang_new_time_step}
\tau_{n+1}=\frac{\tau_n}{\tau_{n-1}}\left(\frac{TOL_t\;r_n}{r_{n+1}\;r_{n+1}}
\right)^{1/3}\tau_n,
\end{equation}
where $TOL_t$ is a desired tolerance prescribed by the user
\cite{GustafssonLundhSoederlind1988}.
If $r_{n+1}>TOL_t$, the step is rejected and redone. Otherwise the step is
accepted and we advance in time.

Observe that the above time discretization scheme has been applied to the abstract Cauchy problem
(\ref{lang_lirk_cauchy}), i.e., to the initial value problem in function space.  This means that the
Rosenbrock discretization scheme (\ref{lang_lirk_rosen1}) is a sequence of linear elliptic
boundary value problems. The spatial approximation of the vectors $U_n$ is now done by
multilevel finite elements. This is described next.

The main idea of multilevel techniques consists of replacing the
infinite dimensional solution space by a nested sequence of finite
dimensional spaces with successively increasing dimension in order to
improve the approximation quality. To construct adaptive spatial meshes,
we apply the edge-oriented hierarchical error estimator from
\cite{DeuflhardLeinenYserentant1989,Lang2000}.
Such estimators are well established to control the adaptive multilevel process producing
successively finer meshes and, in spatial multi-scale cases, drastically
reducing the size of the arising linear algebraic systems with respect to a
prescribed tolerance. Let $\mathcal T_h$ be an admissible finite element mesh at
$t\!=\!t_n$ and $S^1_h$ be the associated finite dimensional space consisting of
all continuous piecewise linear functions. Then the
standard Galerkin finite element approximation $U^h_{ni}\in S^1_h$ of the
intermediate values $U_{ni}$ in (\ref{lang_lirk_rosen1}) satisfies the equation
\begin{equation}
\label{lang_fem_glk}
( L_n\,U^h_{ni} , \phi_h ) = ( R_{ni}, \phi_h )\quad \mbox{ for all }
\phi_h\in S^1_h,
\end{equation}
where $L_n$ is the weak representation of the differential
operator on the left-hand side in (\ref{lang_lirk_rosen1}) and
$R_{ni}$ stands for the entire right-hand side in
(\ref{lang_lirk_rosen1}). Since the operator $L_n$ is independent
of $i$, its calculation is required only once within each time
step. The resulting large scale linear systems are solved by the
BICGSTAB algorithm \cite{vanderVorst1992} with ILU preconditioning.

After computing the approximate intermediate values $U^h_{ni}$, a
posteriori error estimates can be used to give specific assessment
of the error distribution. Consider a hierarchical decomposition
\begin{equation}
\label{lang_lirk_hd}
S^{2}_h = S^1_h \oplus Z^{2}_h,
\end{equation}
where $Z^{2}_h$ is the subspace that corresponds to the span of
quadratic bubble functions corresponding to edges. Defining an
a posteriori error estimator $E^h_{n+1}\in Z^{2}_h$ by
\begin{equation}
\label{lang_fem_femerr} E^h_{n+1} = E^h_{n0} +
\sum\limits_{i=1}^{4}\,m_iE^h_{ni},
\end{equation}
with $E^h_{n0}$ approximating the projection error of the initial value $U_n$
in $Z^{2}_h$ and $E^h_{ni}$ estimating the spatial error of the intermediate
value $U^h_{ni}$, the local spatial error for a finite element $T\in\mathcal T_h$ can
be estimated by $\eta_T\!:=\!\|E^h_{n+1}\|_T$. The error estimator $E^h_{n+1}$
is computed  by linear systems which can be derived from (\ref{lang_fem_glk}).
We get for $i=0$
\begin{equation}
\label{lang_error_init}
(L_n\,E_{n0}^h,\phi_h) = (L_n\,(U_n-U_n^h),\phi_h)\quad\mbox{for all}\; \phi_h\in
Z_h^{2}.
\end{equation}
and for $i=1,\ldots,4$
\begin{equation}
\label{lang_error_stages}
(L_n\,E_{ni}^h,\phi_h) =
(R_{ni} (E_{n1}^h + U_{n1}^h ,\ldots, E_{n,i-1}^h + U_{n,i-1}^h ),\phi_h) -
(L_n\,U_{ni}^h,\phi_h)\quad\mbox{for all}\; \phi_h\in
Z_h^{2}.
\end{equation}
Solving these equations encounters a sequence of five large linear problems in the space of
hierarchical surpluses. From many practical computations, we have experienced
that using the approximate error estimator
\begin{equation}\label{err-simple-def}
E_{n+1}^h \approx \tilde{E}_{n+1}^h = E_{n0}^h+\frac{E_{n1}^h}{\gamma}\,,
\end{equation}
that is an error estimator for the embedded, locally second order linearly implicit
Euler solution $U_{n+1}^{h,euler}=U_n^h+U_{n1}^h/\gamma$, is quite efficient. Combining
(\ref{lang_error_init}) and the first equation of (\ref{lang_error_stages}) yields
the following simplified error equation
\begin{equation}
\label{lang_error_euler}
(L_n\,\tilde{E}_{n+1}^h,\phi_h) =
(L_n(U_n-U_{n+1}^{h,euler})+\frac{R_{n1}}{\gamma},\phi_h)\quad\mbox{for all}\; \phi_h\in
Z_h^{2}.
\end{equation}
Although we have reduced the number of error equations considerably, we still
face a fully coupled system over the surplus space $Z_h^2$ in
\eqref{lang_error_euler}. Following the approach given in \cite{DeuflhardLeinenYserentant1989}, 
we take further advantage of a localization strategy. The idea is to replace the bilinear
form on the left hand side in \eqref{lang_error_euler} by a spectrally
equivalent block-diagonal preconditioner in the surplus space. Then, the error equation can
be simultaneously solved for each bubble function, that is, for each edge in the triangulation.
Let $\tilde{E}_{n+1}^{h,loc}$ be the corresponding error estimator. 
The local spatial error $\eta_T$ for a finite element
$T\in\mathcal T_h$ can again be estimated by computing the norm of
$\tilde{E}_{n+1}^{h,loc}$ over $T$. For the overall spatial error, we define
in line with the local temporal error in (\ref{lang_time_estim})
\begin{equation}\label{lang_space_estim}
|\|\tilde{E}_{n+1}^{h,loc}\||=
\left( \frac{\|\tilde{E}_{n+1}^{h,loc}\|^2_{L^2(Z)}}
{ATOL+RTOL\,\|U^h_{n+1}\|^2_{L^2(Z)}} \right)
^{1/2}.
\end{equation}
Based on this error estimation, we can control the spatial accuracy of the numerically computed 
solution to an imposed tolerance level $TOL_x$. New grid points are placed in 
regions of insufficient accuracy.
Therefore, all elements with $\eta_T>0.8\max_T\eta_T$ are refined. We apply the
standard red-green refinement technique.
The iterative process estimate-refine-solve within a time step is continued until
$|\|\tilde{E}_{n+1}^{h,loc}\||<TOL_x$. Obviously, temporal and spatial errors
have to be well balanced. We have also to take into account mesh coarsening to
gain efficiency. For more details, we refer to \cite{Lang2000}.

\section{Numerical Results}
\label{sec:SIM}
%
\subsection{Marshak Wave}
This test case is a two-dimensional version of the analytical Marshak Wave test case from \cite{SuOls97}. Here, radiation is coupled to an energy equation for $B \sim T^4$. The heat capacity is chosen such that the problem becomes linear. The equations are

\begin{align}
 	 \frac{1}{c}\partial_t \psi(t,x,\Omega) + \Omega \cdot \nabla_x \psi(t,x,\Omega)
	 &= \sigma_a(x)\left(B(t,x)-\psi(t,x,\Omega)\right)+Q(t,x)\label{eq:SuOlsonRHT1}\\
	 \partial_t B(t,x) &= \sigma_a(x)\left(\frac{1}{4\pi}\int_{4\pi} \psi(t,x,\Omega)d\Omega-B(t,x)\right).\label{eq:SuOlsonRHT2}
\end{align}

The $SP_N$ approximation is applied to (\ref{eq:SuOlsonRHT1}) and treats the $B$ variable as an additional source term. The additional equation (\ref{eq:SuOlsonRHT2}) is an ordinary differential equation but fits into the numerical framework above.

The setting is two-dimensional and infinite in space ($x\in\R^2$), with time $t\in[0,10]$. We have $\sigma_a=1.0$. 
In an initially empty medium, a spatially bounded source $Q$ is switched on at time zero:
\begin{equation*}
	Q(t,x) = \begin{cases} \frac{1}{4x_0^2} &\text{for}\quad 0\leq t \leq t^\ast,\ x\in [-x_0,x_0]\times [-x_0,x_0],\\
										0 &\text{otherwise} \end{cases}					
\end{equation*}
with $x_0=0.5$ and $t^\ast=10$. 

\begin{figure}
\subfigure[Cut along $y=0$.]{\epsfig{file=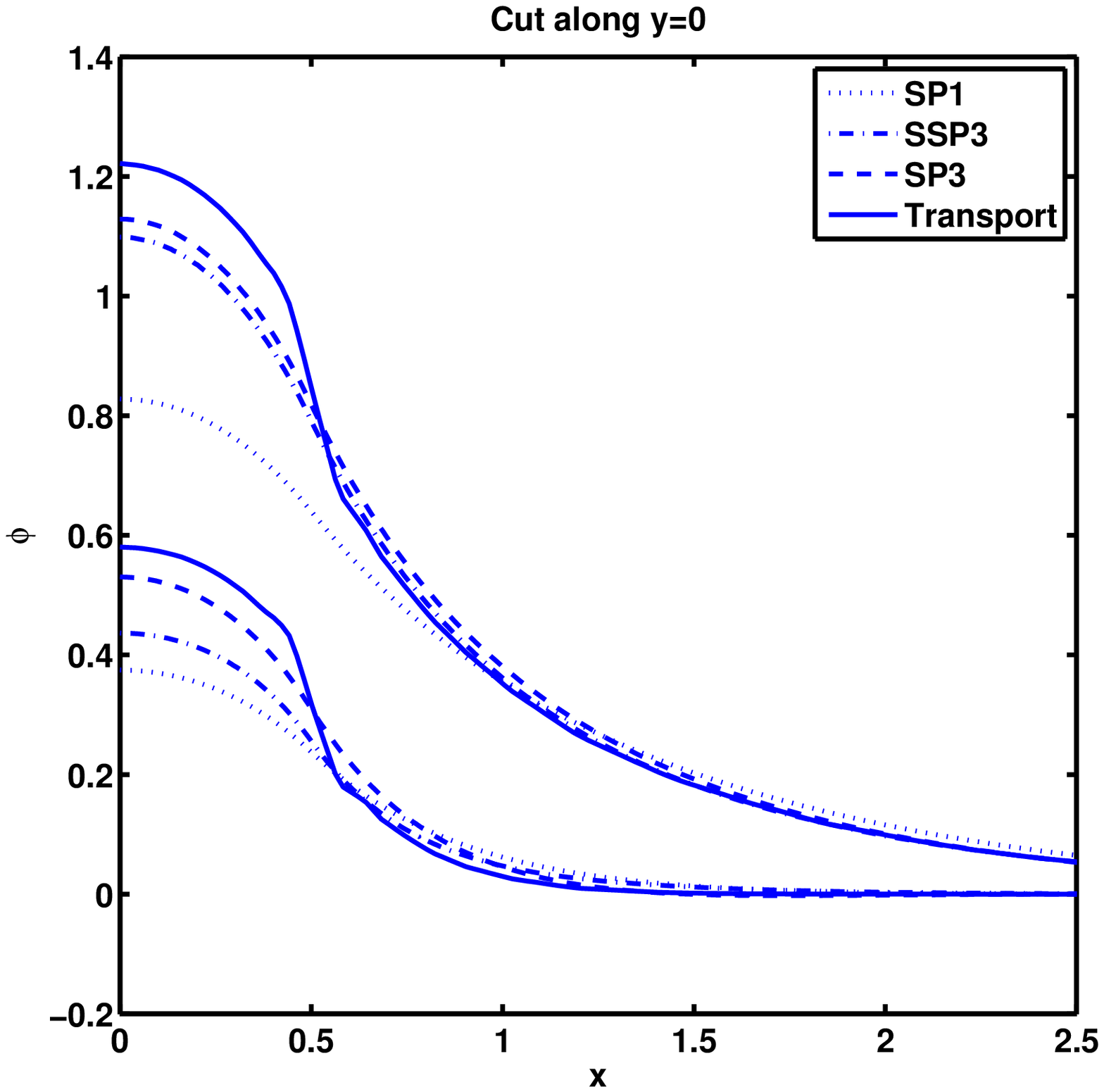,width=0.48\linewidth}\label{fig:Cutx}}
\subfigure[Cut along $x=y$.]{\epsfig{file=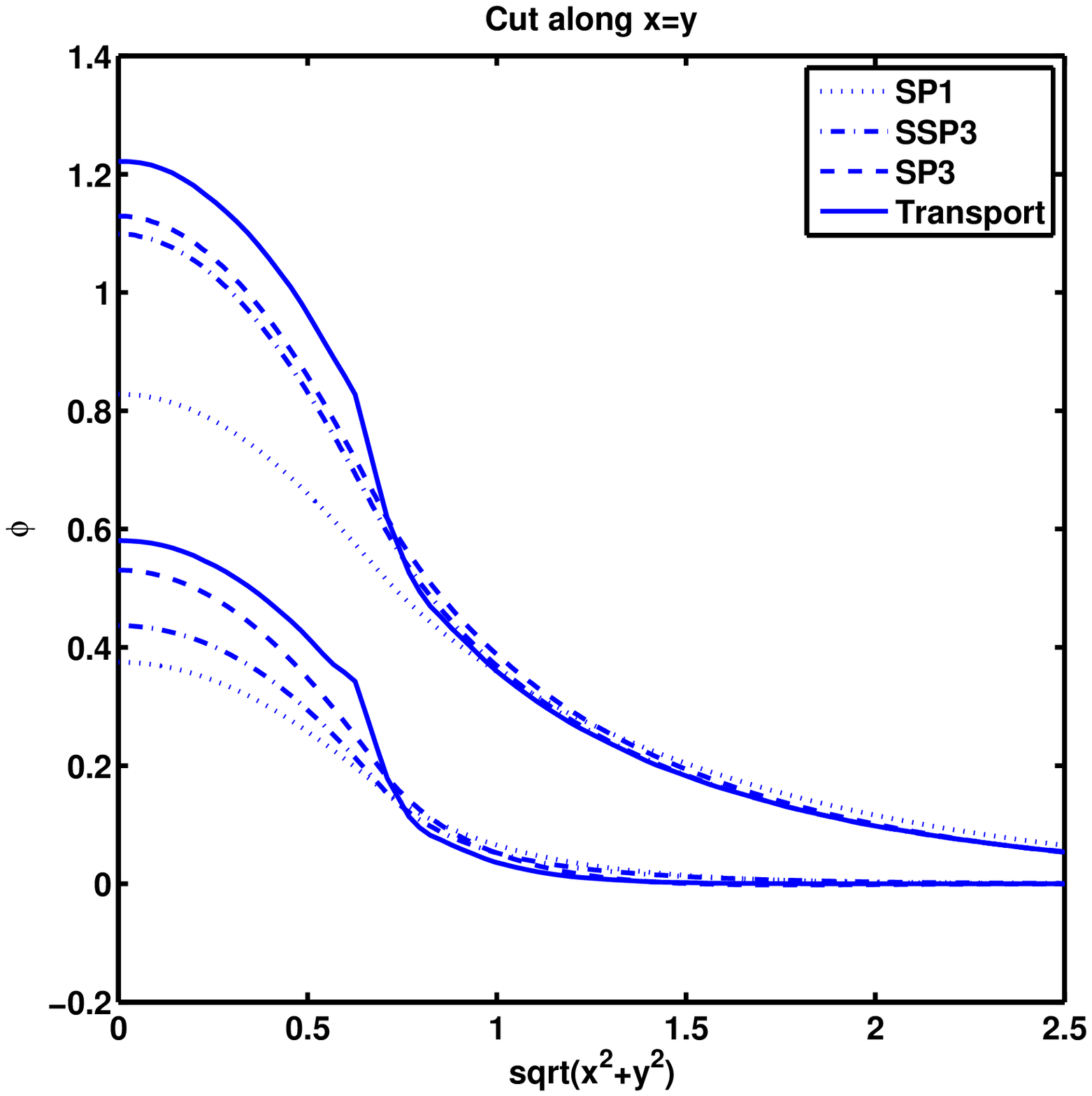,width=0.48\linewidth}\label{fig:Cutdiag}}	
\caption{Energy distribution $\phi$ for different times in 2D Marshak wave.}
\label{fig:2dMarshak}
\end{figure}

\begin{figure}
\begin{center}
\subfigure[Mesh.]{\epsfig{file=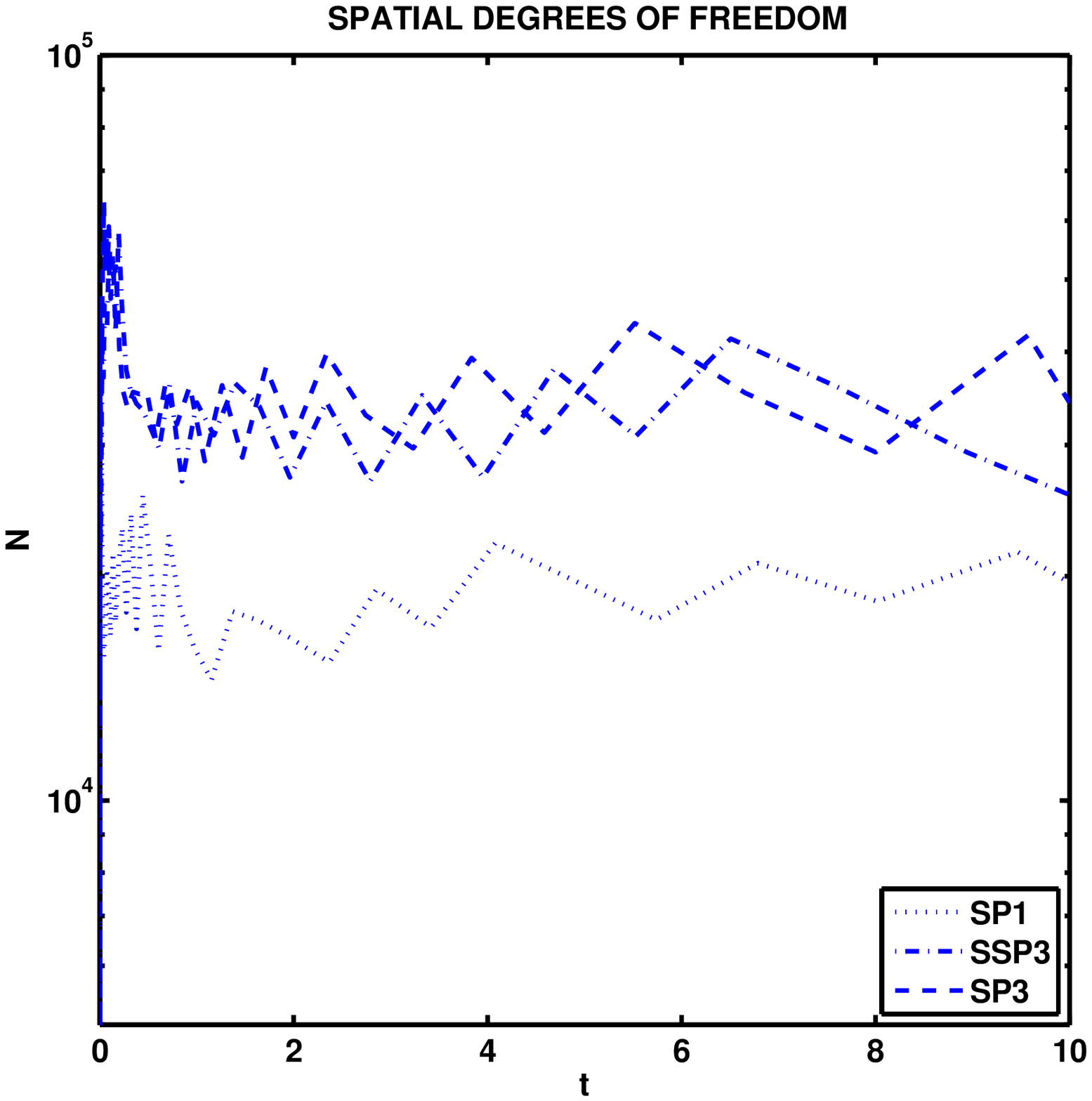,width=0.49\linewidth}\label{fig:marshak-spacedofs}}
\subfigure[Time step.]{\epsfig{file=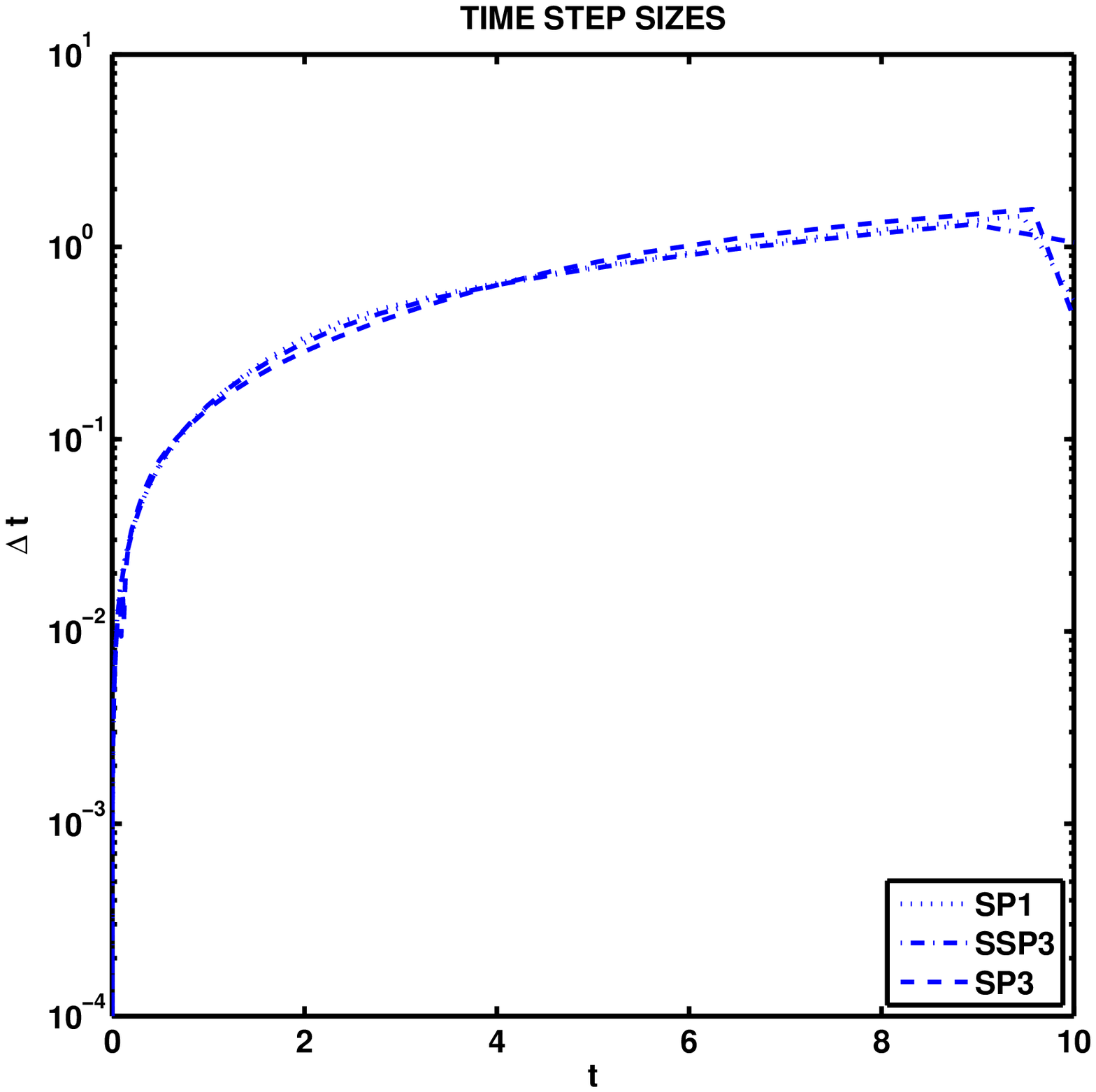,width=0.49\linewidth}\label{fig:marshak-timedofs}}
\caption{Spatial degrees of freedom (left) and step size (right) as functions of time.}
\label{fig:marshak}
\end{center}
\end{figure}

For symmetry reasons, the computational domain was restricted to $[0,10]\times [0,10]$. We set $TOL_t=TOL_x=10^{-4}$, $RTOL=1, ATOL=10^{-4}$. The first time step was $\tau_0=1.0e-4$. We started with a criss-cross grid of $20^2 + 21^2= 841$ points, the largest side length in one triangle being $h=2^{-1}$. After adaptive refinement, the smallest side length was $h = 2^{-10}\sqrt{2}$. The finest grid consisted of 58916 points (both values extremal for $SP_3$). Both spatial degrees of freedom and the time step are shown in Figure \ref{fig:marshak}.
The total computation times were 19m33s ($SP_1$), 79m10s ($SSP_3$), 208m53s ($SP_3$) on a PC with a 3 GHz i686 processor.

In Figure \ref{fig:2dMarshak}, the radiative energy $\phi$ for times $t=1$ (lower curves) and $t=10$ (higher curves) computed by the different models $SP_1$, $SP_3$, $SSP_3$ is compared to a benchmark solution (high-order spherical harmonics solution), for both a cut along the $x$ axis as well as a cut along the diagonal $x=y$. 
As in the one-dimensional case (investigated in \cite{FraKlaLarYas07}, the higher-order diffusion approximations are a clear improvement on the $SP_1$ diffusion approximation. 




\subsection{Lattice Problem}
\begin{figure}
\begin{center}
\subfigure[Mesh.]{\epsfig{file=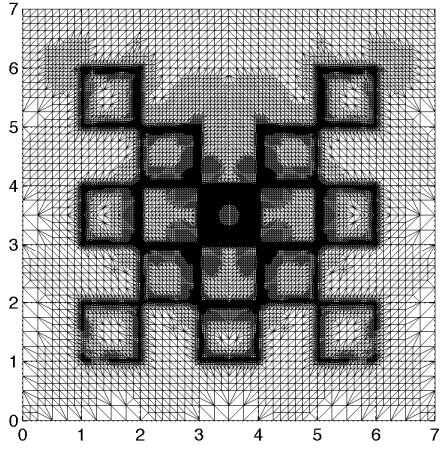,width=0.55\linewidth}\label{fig:mesh}}
\subfigure[Time step.]{\epsfig{file=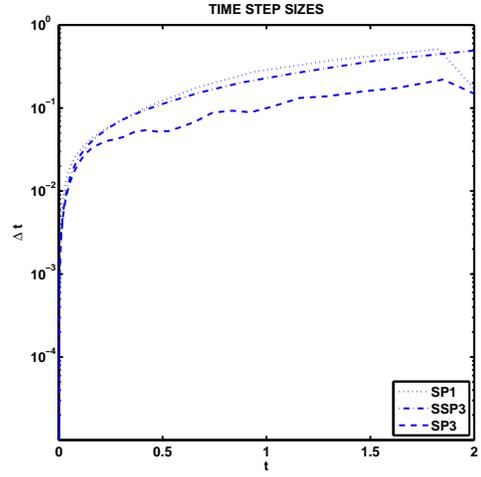,width=0.44\linewidth}\label{fig:lattice-timedofs}}
\caption{Computational mesh for the $SP_3$ solution at $t=2.0$ (left) and step size as a function of time (right).}
\label{fig:lattice}
\end{center}
\end{figure}

\begin{figure}
\begin{center}
\subfigure[Setting.]{\epsfig{file=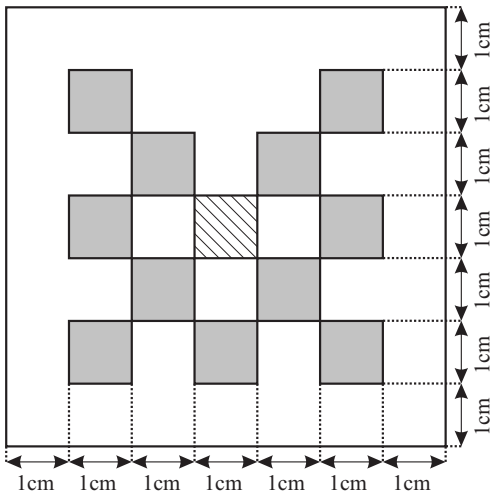,width=0.4\linewidth}\label{fig:BruHolGeometry}}\hfill
\subfigure[Cut along $x=3.5$.]{\epsfig{file=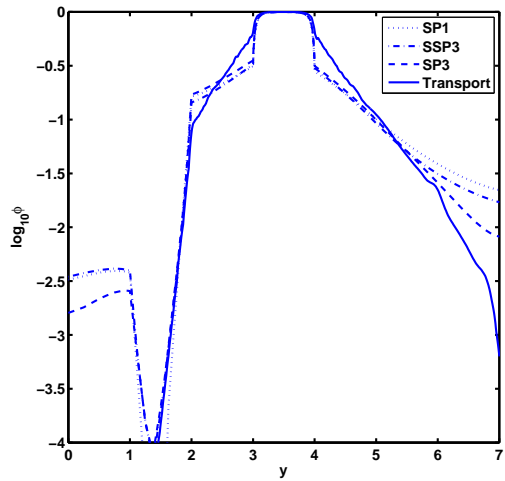,width=0.55\linewidth}\label{fig:DiagCut}}	
\caption{Computational domain for the $SP_3$ solution (left) and comparison of the different radiative energies along the center line $x=3.5$.}
\end{center}
\end{figure}

As a more complex numerical test we consider a 2D checkerboard structure of different
materials. The geometry, shown in Fig.~\ref{fig:lattice}, is identical to
the example presented in \cite{BruHol05}, however modified here to have
$\sigma_t = \sigma_s =0.2$ cm$^{-1}$ in the highly scattering regions
(white in Fig.~\ref{fig:lattice}), and
$\sigma_t=10$ cm$^{-1}$, $\sigma_s=0$ cm$^{-1}$ in the highly absorbing regions
(grey and hatched squares in Fig.~\ref{fig:BruHolGeometry}).
A source (hatched square in Fig.~\ref{fig:lattice}) $q=1$ is switched on at
$t=0$. The final time is $t=2.0$. The test lies in an intermediate regime between thin
and thick media. The radiation field propagates through several obstacles. The main front propagates to the top which is open, but some radiation will leak through the squares.

\begin{figure}
\subfigure[$P_7$ solution.]{\epsfig{file=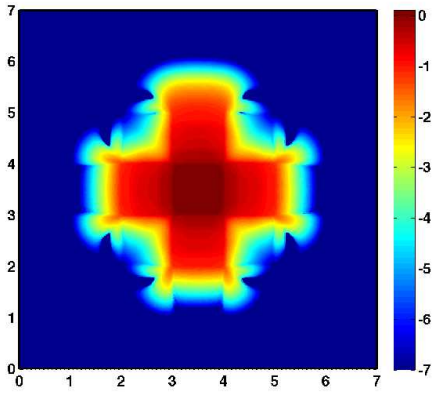,width=0.48\linewidth}}
\subfigure[$SP_1$ solution.]{\epsfig{file=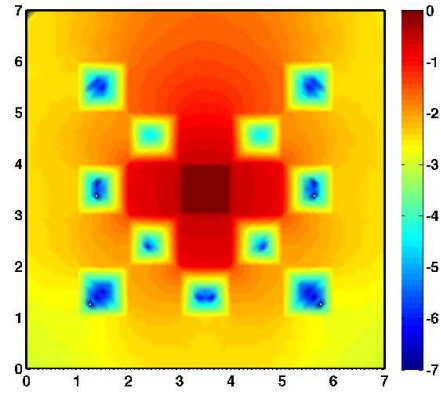,width=0.48\linewidth}}\\
\subfigure[$SSP_3$ solution.]{\epsfig{file=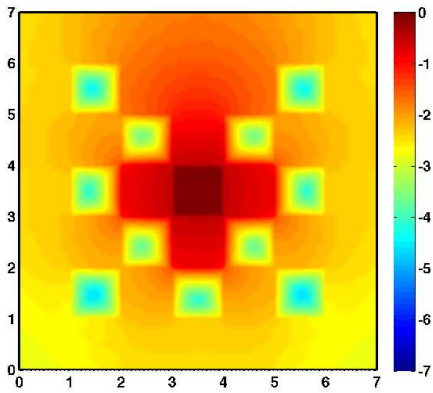,width=0.48\linewidth}}
\subfigure[$SP_3$ solution.]{\epsfig{file=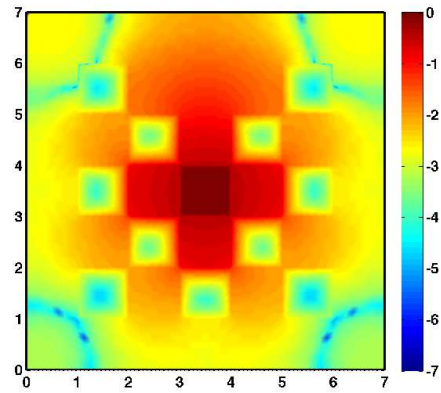,width=0.48\linewidth}}
\caption{Energy distribution at $t=2.0$ for different models.}
\label{fig:LatticeResult}
\end{figure}

We set $TOL_t=10^{-3}, TOL_x=10^{-4}$, $RTOL=1$, $ATOL=10^{-7}$ with initial time step $\tau_0=10^{-5}$. Figure \ref{fig:mesh} shows the adaptive grid for the $SP_3$ solution at $t=2.0$. It consists of 100100 triangles and 50135 points, the smallest triangle being $2^{-9}$. Local refinements can be seen around the jumps at the interfaces between the two media and in the central radiating region. Computation times were 23m46s ($SP_1$), 58m16s ($SPP_3$), 164m57s ($SP_3$) on a PC with a 3 GHz i686 processor.

Figure \ref{fig:LatticeResult} shows the radiative energy at time $t=2.0$ on a logarithmic scale. The $P_7$ solution, which serves as a benchmark here, shows sharp fronts filling the squares adjacent to the center square and a small front escaping on the top. While the radiation in the center region is sufficiently well-described by the diffusion solutions, they overpredict the spreading of the front into the outer regions. However, the $SP_3$ solution clearly has a smaller and sharper front than the other two lower-order approximations. Quantitatively, this becomes more clear when looking at a cut through the energy profile at $x=3.5$, shown in Figure \ref{fig:DiagCut}. The escaping front to the top is between $y=5$ and $y=7$ and it becomes clear that $SP_3$ is closest to the benchmark, with $SSP_3$ having slight advantages over $SP_1$.




%

\section{Conclusions}
\label{sec:CON}
Concerning the validity of the time-dependent $SP_N$ equations, assertions similar to the steady case can be made. Physically, the parabolic scaling and $\eps$ small mean that we require the ratio of mean free path and characteristic length scale, as well as the characteristic length divided by the product of characteristic time and velocity to be small of order $\eps$. 

The numerical results here and in \cite{FraKlaLarYas07} indicate that the $SP_N$ approximations improve diffusion theory in the sense that not too far away from the diffusive limit a better approximation is obtained.  

In the lattice test case, a strong grid refinement around the obstacles was necessary. Similarly, to approximate the effect of source terms well, initially smaller time steps were needed. Using adaptivity in both time and space discretization is essential in many radiative transfer applications. This is especially true for non-homogeneous media with varying coefficients or with source terms.

\section*{Acknowledgments}
%
The authors acknowledge support from the German Research Foundation DFG under grant KL 1105/14/2 and SFB 568/3.

\bibliographystyle{amsplain}
\bibliography{RadLit,RadLitLang}

\end{document}